\documentclass{article}
\usepackage{tikz}
\usepackage{cite}  
\def\checkmark{\tikz\fill[scale=0.4](0,.35) -- (.25,0) -- (1,.7) -- (.25,.15) -- cycle;} 
\usepackage{booktabs}
\usepackage{graphicx}
\usepackage{longtable}
\usepackage{hyperref}
\usepackage{multirow}
\usepackage{amsmath}
\usepackage{adjustbox}

\begin{document}

\title{ML-Dev-Bench: Comparative Analysis of AI Agents on ML development workflows}
\author{Harshith Padigela\textsuperscript{*} \and Chintan Shah\textsuperscript{*} \and Dinkar Juyal\textsuperscript{*}}
\footnotetext{\textsuperscript{*}Equal contribution. Author Correspondance: harshith2794@gmail.com}
\date{}
\maketitle

\begin{table}[t]
\centering
\begin{tabular}{|l|p{5.5cm}|}
\hline
\textbf{Category} & \textbf{Description} \\
\hline
Dataset Handling & Downloading and preprocessing datasets \\
\hline
Model Training & Loading pretrained models, fine-tuning \\
\hline
Debugging & Addressing errors in training files, exploding gradients, and incorrect implementations \\
\hline
Model Implementation & Modifying and implementing on top of existing model architectures \\
\hline
API Integration & Integrating logging tools like WandB \\
\hline
Performance & Improving baselines and achieving competitive results \\
\hline
\end{tabular}
\caption{Task Categories and Their Descriptions}
\label{tab:task_categories}
\end{table}

\section{Abstract}
In this report, we present ML-Dev-Bench\footnote{\url{https://github.com/ml-dev-bench/ml-dev-bench}}, a benchmark aimed at testing agentic capabilities on applied Machine Learning development tasks. While existing benchmarks focus on isolated coding tasks or Kaggle-style competitions, ML-Dev-Bench tests agents' ability to handle the full complexity of ML development workflows. The benchmark assesses performance across critical aspects including dataset handling, model training, improving existing models, debugging, and API integration with popular ML tools. We evaluate three agents - \textbf{ReAct}, \textbf{Openhands}, and \textbf{AIDE} - on a diverse set of 30 tasks, providing insights into their strengths and limitations in handling practical ML development challenges.
We open source the benchmark for the benefit of the community at \href{https://github.com/ml-dev-bench/ml-dev-bench}{https://github.com/ml-dev-bench/ml-dev-bench}.

\section{Introduction}
Recent advances in Large Language Models (LLMs) have demonstrated impressive capabilities in code generation and software engineering tasks. This has led to the development of various benchmarks like HumanEval \cite{chan2024mlebenchevaluatingmachinelearning}, MBPP  \cite{austin2021programsynthesislargelanguage} that evaluate coding abilities, and others like SWE-Bench \cite{jimenez2024swebenchlanguagemodelsresolve}, that test LLM-based agents on software engineering tasks. However, while these benchmarks effectively assess general programming capabilities, they don't capture the unique challenges of Machine Learning development workflows, 

Benchmarks such as ML-Bench \cite{tang2024mlbenchevaluatinglargelanguage}, test agents’ abilities to generate code and commands to interact with popular ML repositories, while MLE-Bench \cite{chan2024mlebenchevaluatingmachinelearning} and MLAgentBench \cite{huang2024mlagentbenchevaluatinglanguageagents} focus on Kaggle-style tasks to evaluate the iterative and open-ended nature of ML development. However, real-world ML development extends far beyond that, including the complexity of working on top of existing codebases and models, integrating with third-party tools, debugging complex issues that span multiple components of the ML pipeline, and understanding and balancing trade-offs like model performance and cost to come up with optimal design. 

ML-Dev-Bench addresses this gap by providing a comprehensive evaluation framework that tests an agent's ability to handle real-world ML development scenarios. Our benchmark is particularly relevant,, as ML development increasingly relies on large language models and AI agents to assist developers. Understanding the capabilities and limitations of these agents in handling practical ML development tasks is crucial for their effective deployment in production environments.

\section{Benchmark Design}

ML-Dev-Bench comprises of 30 carefully designed tasks that evaluate various aspects of ML development. These tasks are structured to assess both specific technical capabilities (like handling datasets, model implementation) and broader problem-solving skills (like model training and performance improvement) that are essential in real-world ML development.
The tasks span several key categories of ML development shown in Table \ref{tab:task_categories}:

\begin{enumerate}
    \item Dataset Handling focuses on evaluating the ability to work with large datasets, inspect them and apply pre-processing pipelines. An example is the noisy imagenette \cite{imagenette2019} dataset download task, where the agent needs to download the dataset, inspect its contents to identify the labels file, only load the 50\% noisy labels from it and generate class summary statistics.  
    \item Model Training tests an agent's ability to work with existing models, from loading pretrained weights to implementing training loops, logging metrics and managing the training process. These tasks assess both technical skills and the ability to handle long-running tasks.
    \item Debugging presents common scenarios including shape errors, exploding gradients, incorrect implementations, and integration errors. Agents must analyze large training logs, metrics, and code across multiple files to identify and resolve issues.
    \item Model Implementation tests the ability to modify existing architectures and implement new features. An example is the ChannelViT related tasks, which follow three levels of increasing difficulty:
Level 1 provides complete specifications with examples and tests;
Level 2 includes specifications and tests but omits examples;
Level 3 gives specifications but tests and examples are hidden
    \item API Integration assesses the ability to work with essential ML development tools, particularly for logging and experiment tracking.
    \item Model Performance tasks challenge agents to improve baseline implementations through iterative experimentation and hypothesis testing.
\end{enumerate}

\begin{table}[t]
\small
\setlength{\tabcolsep}{4pt}
\centering
\renewcommand{\arraystretch}{1.2}
\begin{tabular}{|l|c|c|c|c|c|}
\hline
\textbf{Category} & \textbf{ReAct-Sonnet} & \textbf{OH-Sonnet} & \textbf{OH-Gemini}  & \textbf{Aide-4o} & \textbf{ReAct-4o}  \\
\hline
Dataset Handling & 100\% (3/3) & 100\% (3/3) & 66\% (2/3) &  33\% (1/3) & 0\% (0/3) \\
API Integration & 100\% (1/1) & 100\% (1/1) &  0\% (0/1) & 0\% (0/1) & 100\% (1/1) \\
Model Training & 67\% (4/6) & 83\% (5/6) & 33\% (2/6) & 33\% (2/6) & 50\% (3/6) \\
Debugging & 57\% (4/7) & 57\% (4/7) & 14\% (1/7) & 29\% (2/7) & 14\% (1/7) \\
Model Implementation & 29\% (2/7) & 29\% (2/7) & 0\% (0/7) & 0\% (0/7) & 0\% (0/7) \\
Model Performance & 0\% (0/6) & 0\% (0/6) & 0\% (0/6) & 0\% (0/6) & 0\% (0/6) \\
\hline
\textbf{Overall} & \textbf{47\% (14/30)} & \textbf{50\% (15/30)} & \textbf{17\% (5/30)} & \textbf{17\% (5/30)} & \textbf{17\% (5/30)} \\
\hline
\end{tabular}
\caption{Category-wise Success Rates Across AI Agents with different models}
\label{tab:category_success_rates}

\end{table}

\subsection{Evaluation Metrics}

Tasks are evaluated based on binary success (\checkmark) or failure (\texttimes). The aggregate success rate for each agent is calculated as:

\begin{equation}
\text{Success Rate} = \frac{\text{Total Successful Tasks}}{\text{Total Tasks}} \times 100\%
\end{equation}

Agents are assessed on their ability to complete tasks accurately without introducing errors or hallucinations.

\section{Evaluation Framework}
In this section we briefly describe the design of our evaluation framework, called Calipers, for running the benchmark. 
The framework consists of three components: agents, evaluation tasks, and metrics. Agents are evaluated on various Machine Learning tasks to generate metrics.
We designed Calipers to allow easy addition of new evaluation tasks, agents and metrics, ensuring the benchmark can evolve alongside advances in ML development practices and tooling.

\subsection{Evaluation Task}
Each evaluation task consists of a task description, a set of input code and data files, and a validation logic that checks the correctness of the outputs and artifacts generated by the agent.
Depending on the type of task, we implement various types of validation checks including
\begin{itemize}
    \item Running tests on generated code to check for correctness
    \item Checking for the presence of all required output files and artifacts
    \item Evaluating agent generated model checkpoints for required performance
    \item Querying logged artifacts and metrics from wandb
\end{itemize}

\subsection{Agents}

Each agent is provided with two inputs in an evaluation run, the description of the task and a working directory populated with initial input files.
The agent's outputs are task-specific artifacts which are saved in the working directory. These outputs are validated to determine success or failure. We generate the evaluation metrics discussed in the previous section for each evaluation run. We use litellm callbacks to capture metrics like number of steps, tokens, and cost.

\section{Agent Setup}
We evaluate three agents on ML-Dev-bench. The agents and their setup is described below. 
Each agent uses an LLM and a set of tools to execute various actions. All agents execute their code in a runtime environment which is either a local python or docker environment depending on the agent. We customized the runtime environments for all agents to pre-install common ML frameworks like scikit-learn, pytorch, transformers, lightning, wandb, etc to ensure smooth execution.

\begin{enumerate}
    \item \textbf{ReAct:} We created a simple ReAct agent \cite{yao2023reactsynergizingreasoningacting} as a baseline which takes actions by calling tools. We used the \href{https://www.langchain.com/langgraph}{LangGraph} framework for the agent and \href{https://github.com/ComposioHQ/composio}{Composio} toolset which provides tools for common use cases. We customized the tools to reliably capture large command outputs, handle long running commands and ensure consistency across different tools like file and shell tools. All the tool calls were executed in a local python environment which was pre-installed with common ML frameworks as mentioned earlier and had access to the relevant api keys. No custom prompts were used, and the agent was allowed to run for a maximum of 50 steps. 
    We tested the agent with Claude Sonnet 3.5 10-2022 and OpenAI GPT-4o.
    \begin{enumerate}
        \item \textbf{Command line tools}
        \begin{enumerate}
            \item \textbf{Shell Tool} - to execute short running commands
            \item \textbf{Spawn Tool} - to execute long running commands like training in the background
            \item \textbf{Sleep and execute tool} - to wait and monitor long running processes
        \end{enumerate}
        \item \textbf{File tools} like create files, list files and edit files
    \end{enumerate}
    \item \textbf{Openhands:} Openhands \cite{wang2024openhandsopenplatformai} is a popular open-source coding agent with state-of-the-art performance on SWE-Bench-Full \cite{jimenez2024swebenchlanguagemodelsresolve}. We used Openhands agent v0.21.1 and customized the runtime build to install common ML frameworks listed above. We tested the agent with Claude Sonnet 3.5 10-2022 model which is the current best performing model with the agent on SWE Bench and Gemini 2.0 Flash. The agent was allowed to run for a maximum of 50 steps.
    \item \textbf{AIDE:} \href{https://github.com/WecoAI/aideml}{AIDE} is an agent purpose-built for data science tasks like Kaggle competitions \cite{chan2024mlebenchevaluatingmachinelearning} and performs a tree search over solutions. AIDE scaffolding performs better in comparison to other agents like Openhands on MLEBench using o1, GPT-4o. Unlike other general purpose agents which output any artifact, AIDE outputs an evaluation metric and code as its final output. All other artifacts are considered intermediate outputs and saved in a custom working directory. Since not all tasks in ML-Dev-Bench require outputting a score, we access the artifacts from its custom working directory to validate the agent's performance. Given the high costs of o1, we evaluated the agent with GPT-4o.

\end{enumerate}

\section{Performance Comparison}
Performance of the agents across different task categories, Table \ref{tab:category_success_rates} and individual tasks Table \ref{tab:performance_comparison} reveals a consistent pattern. Performance decreases as tasks become more open-ended and complex. The success rates are highest in well-defined categories like dataset handling and basic debugging with clear instructions, but drop significantly in open-ended and long-running tasks like model performance tasks where no agent succeeded.

OpenHands-Sonnet (OH-Sonnet) and ReAct-Sonnet are the two best performing agents performing agents with 50\% and 47\%
success rate respectively, while OH-Gemini, AIDE-4o and React-4o achieve 17\% success rate.

Across the 14 common successful tasks between OH-Sonnet and ReAct-Sonnet, we did not observe any discernible trends in cost. In long running tasks, OH-Sonnet used more tokens on average, especially in debugging and training tasks, while ReAct had higher usage in model implementation tasks like ChannelViT. Even in tasks with higher token usage, the costs dont scale proportionally due to efficient prompt caching in Openhands.

\subsection{ReAct-Sonnet}

ReAct-Sonnet had a success rate of 47\% (14/30 tasks), demonstrating good performance in specific, well-defined tasks like dataset handling (3/3 tasks), basic model training (4/6 tasks), and debugging tasks with clear specifications (4/7 tasks). However, its performance degraded significantly in model implementation tasks where it needed to modify multiple files and in model performance tasks which are open-ended and long-running. The agent had some common failure modes we list below:
\begin{enumerate}
    \item Excessive verification seeking: The agent has a tendency to request feedback even when its instructed to run till completion, specially in complex tasks like model implementation and training.
    \item Premature task termination: In long-running training scenarios, it doesnt wait until completion of tasks but returns control saying training should complete and achieve required performance.
    \item In long running tasks it fails to successfully implement sub-tasks which it handled correctly in isolation. For example, it correctly handles the noisy imagenette dataset setup in dedicated download tasks but fails the same operation when it's part of the longer training pipeline.
\end{enumerate}

The agent's token usage and costs varied significantly across tasks. Simple operations like dataset downloads cost around $0.02-0.08\$$, while debugging tasks cost between $0.1-0.4\$$, some tasks like ChannelViT-Easy debugging take more steps indicating potentially inefficient exploration in complex scenarios.

\subsection{OpenHands-Sonnet}
OpenHands-Sonnet demonstrated the highest success rate at 50\% (15/30 tasks), showing robust performance across most categories. The agent successfully completed all dataset handling tasks (3/3) and showed strong performance in model training (5/6) and debugging (4/7). 

The agent particularly excelled in structured tasks and showed better persistence in long-running operations compared to ReAct-Sonnet. However, it still failed to complete any of the model performance tasks (0/5), indicating limitations in open-ended problem-solving scenarios requiring iterative improvement.

\subsection{OpenHands-Gemini}

OpenHands-Gemini achieved a 17\% success rate (5/30 tasks), demonstrating limited capabilities across most evaluation categories. The agent showed some competence in basic dataset handling tasks and setting up pretrained models for training purposes. However, it performed suboptimally in debugging (1/7), model implementation (0/7), and performance improvement tasks (0/6).
The agent exhibited several consistent failure modes. It frequently failed to follow task instructions regarding creation of output artifacts. In debugging scenarios, it often modified test files despite explicit instructions not to do so. The agent also demonstrated significant limitations in file editing capabilities and fixing code errors.
\subsection{ReAct-4o}
ReAct-4o with 17\% success rate  (5/30 tasks) had some success with tasks with well-defined specifications (WandB logging, downloading a specific model from Torchvision), and certain debugging tasks. However it did struggle on other tasks in the same categories. It also failed on the relatively easier tasks like dataset download due to not following instructions, ran into indentation errors while attempting to debug code and failed to produce output artifacts as required by certain tasks.

\subsection{Aide-4o}
Aide-4o had a 17\% success rate (5/30 tasks), demonstrating limitations across most categories. The agent managed to complete some basic dataset handling and debugging tasks but struggled with model training (2/6) and completely failed in model implementation and model performance categories.

The cost metrics for Aide-4o weren't captured as it doesnt use LiteLLM, but its low success rate across both specific and open-ended tasks suggests limitations in handling ML development workflows. The agent's successes were primarily limited to tasks with very clear, step-by-step instructions and immediate feedback loops.

\section{Conclusion}
We presented ML-Dev-Bench, a benchmark focused on ML development workflows consisting of 30 tasks. We evaluated 3 agents on this benchmark - ReAct (with Claude Sonnet and GPT-4o), Openhands and AIDE; Openhands with Claude Sonnet performed the best out of these. Our open-source evaluation framework provides a foundation for the community to build upon. We encourage researchers and practitioners to explore various directions: analyzing the impact of scaling compute on these agents, studying variance in success metrics across multiple runs, evaluating emerging reasoning models like DeepSeek-R1 and o1/o3, and expanding the problem categories to include areas such as label and data collection. We welcome contributions to enhance the benchmark's scope, methodology, and agent evaluation pipeline.
\bibliographystyle{plain}  
\bibliography{references}  

\newpage
\section{Appendix}
We share the traces of agentic runs for different tasks in Google Drive here \url{https://drive.google.com/drive/folders/1o1FCvx_n9XVKgvkSWL97LlsfOh_KfG-S?usp=sharing}.

We also share the comparison of token metrics across agentic frameworks while keeping the base model constant, and a detailed breakdown of performance at a per-task level below.

\begin{table}[t]
\centering
\renewcommand{\arraystretch}{0.9} 
\begin{tabular}{|p{3.5cm}|p{3.5cm}|p{1.0cm}|p{1.0cm}|p{1.0cm}|p{1.0cm}|p{1.6cm}|}
\hline
\textbf{Task} & \textbf{Category} & \textbf{ReAct-Sonnet} & \textbf{OH-Sonnet} & \textbf{Aide-4o} & \textbf{ReAct-4o} & \textbf{OH-Gemini-2.0-Flash} \\
\hline
Dataset download - Noisy Imagenette & Dataset Handling & \checkmark & \checkmark & \texttimes & \texttimes & \texttimes \\
Dataset download - dataset does not exist & Dataset Handling & \checkmark & \checkmark & \checkmark & \texttimes & \checkmark \\
Dataset preprocessing & Dataset Handling & \checkmark & \checkmark & \texttimes & \texttimes & \checkmark \\
Pretrained model download - Torchvision & Model Training & \checkmark & \checkmark & \checkmark & \checkmark & \checkmark \\
Pretrained model download - HuggingFace & Model Training & \checkmark & \checkmark & \texttimes & \texttimes & \checkmark \\
Vision finetuning - classification & Model Training & \checkmark & \checkmark & \texttimes & \texttimes & \texttimes \\
Overfit on small dataset & Model Training & \checkmark & \checkmark & \checkmark & \checkmark & \texttimes \\
Large training logs & Model Training & \texttimes & \checkmark & \texttimes & \checkmark & \texttimes \\
CIFAR10 Training & Model Training & \texttimes & \texttimes & \texttimes & \texttimes & \texttimes \\
Fix problems in model and dataloader & Debugging & \texttimes & \texttimes & \texttimes & \texttimes & \texttimes \\
Model forward pass - shape mismatches & Debugging & \checkmark & \checkmark & \checkmark & \checkmark & \texttimes \\
Model Training - shape mismatches & Debugging & \checkmark & \checkmark & \checkmark & \texttimes & \texttimes \\
NaN losses & Debugging & \checkmark & \checkmark & \texttimes & \texttimes & \checkmark \\
Correct norm for pretrained model & Debugging & \checkmark & \checkmark & \texttimes & \texttimes & \texttimes \\
TinyBERT Eval & Debugging & \texttimes & \texttimes & \texttimes & \texttimes & \texttimes \\
ViT debugging & Debugging & \texttimes & \texttimes & \texttimes & \texttimes & \texttimes \\
Wandb integration & API Integration & \checkmark & \checkmark & \texttimes & \checkmark & \texttimes \\
ChannelViT - Easy & Model Implementation & \checkmark & \checkmark & \texttimes & \texttimes & \texttimes \\
ChannelViT & Model Implementation & \checkmark & \checkmark & \texttimes & \texttimes & \texttimes \\
ChannelViT - No tests & Model Implementation & \texttimes & \texttimes & \texttimes & \texttimes & \texttimes \\
VAR implementation & Model Implementation & \texttimes & \texttimes & \texttimes & \texttimes & \texttimes \\
Multi-head Latent Attention & Model Implementation  & \texttimes & \texttimes & \texttimes & \texttimes & \texttimes \\
Multi-head Latent Attention - hidden tests & Model Implementation  & \texttimes & \texttimes & \texttimes & \texttimes & \texttimes \\
Proximal Policy Optimization & Model Implementation & \texttimes & \texttimes & \texttimes & \texttimes & \texttimes \\
Improve CIFAR-10 baseline - existing model ckpt & Performance & \texttimes & \texttimes & \texttimes & \texttimes & \texttimes \\
Noisy Imagenette & Performance & \texttimes & \texttimes & \texttimes & \texttimes & \texttimes \\
CIFAR-10 long tailed & Performance & \texttimes & \texttimes & \texttimes & \texttimes & \texttimes \\
Segmentation & Performance & \texttimes & \texttimes & \texttimes & \texttimes & \texttimes \\
BoolQ & Performance & \texttimes & \texttimes & \texttimes & \texttimes & \texttimes \\
CIFAR-100 baseline improvement & Performance & \texttimes & \texttimes & \texttimes & \texttimes & \texttimes \\
\hline
\textbf{Success Rate} &  & \textbf{47\% (14/30)} & \textbf{50\% (15/30)} & \textbf{17\% (5/30)} & \textbf{17\% (5/30)} & \textbf{17\% (5/30)} \\
\hline
\end{tabular}
\caption{Performance Comparison Across AI Agents}
\label{tab:performance_comparison}
\end{table}

\begin{table}[t]
\begin{adjustbox}{margin=0pt 0pt 0pt -20pt}  

\centering
\renewcommand{\arraystretch}{1.2} 
\setlength{\tabcolsep}{5pt} 
\begin{tabular}{|p{7.5cm}|r|r|r|r|}
\hline
{\textbf{Task}} & \multicolumn{2}{c|}{\textbf{Token Cost (\$)}} & \multicolumn{2}{c|}{\textbf{Total Tokens}} \\
\cline{2-5}
& \textbf{ReAct} & \textbf{OH} & \textbf{ReAct} & \textbf{OH} \\
\hline
Vision finetuning - classification & 0.176 & 0.124 & 67,034 & 54,434 \\
ChannelViT & 1.06 & 0.215 & 338,055 & 177,182 \\
ChannelViT - Easy & 1.090 & 0.318 & 352,141 & 267,508 \\
ChannelViT - No tests & 0.091 & 0.121 & 33,208 & 53,475 \\
Dataset download - dataset does not exist & 0.018 & 0.051 & 13,735 & 16,125 \\
Dataset preprocessing & 0.078 & 0.103 & 27,210 & 42,277 \\
Model forward pass - shape mismatches & 0.069 & 0.075 & 27,629 & 44,396 \\
Pretrained model download - HuggingFace & 0.096 & 0.063 & 36,592 & 24,738 \\
CIFAR-10 long tailed & 0.089 & 0.334 & 29,659 & 351,943 \\
Fix problems in model and dataloader & 0.376 & 0.556 & 146,826 & 785,182 \\
NaN losses & 0.124 & 0.212 & 40,385 & 193,801 \\
Dataset download - Noisy Imagenette & 0.129 & 0.068 & 67,426 & 25,429 \\
Correct norm for pretrained model & 0.380 & 0.265 & 136,831 & 313,149 \\
Overfit on small dataset & 0.093 & 0.133 & 25,642 & 52,791 \\
Large training logs & 0.023 & 0.185 & 35,189 & 114,903 \\
Segmentation & 0.118 & 0.383 & 39,662 & 395,127 \\
Pretrained model download - Torchvision & 0.058 & 0.044 & 22,289 & 30,596 \\
CIFAR10 Training & 0.209 & 0.288 & 71,409 & 253,445 \\
Model Training - shape mismatches & 0.289 & 0.147 & 115,568 & 92,262 \\
Add implementation - VAR & 0.051 & 0.139 & 12,473 & 59,863 \\
Wandb integration & 0.155 & 0.258 & 65,810 & 266,738 \\
TinyBERT Eval & 0.313 & 0.573 & 121,773 & 762,780 \\
ViT debugging & 0.277 & 1.496 & 84706 & 2,526,426 \\
BoolQ & 0.343 & 0.650 & 131,458 & 993,175 \\
Improve CIFAR-10 baseline - existing model ckpt & 0.115 & 0.401 & 29,232 & 366,660 \\
Noisy Imagenette & 0.582 & 0.288 & 192,275 & 253,445 \\
Multi-head Latent Attention & 3.164 & 1.910 & 1,022,125 & 1,706,146 \\
Multi-head Latent Attention - no hidden tests & 0.160 & 1.278 & 40,068 & 380,506 \\
Proximal Policy Optimization & 0.369 & 3.048 & 87,988 & 904,267 \\
CIFAR-100 baseline improvement & 0.123 & 3.531 & 30,493 & 1,133,630 \\

\hline
\end{tabular}
\label{tab:token-metrics}
\end{adjustbox}
\caption{Comparison of Token Metrics between ReAct-Sonnet and OpenHands-Sonnet}

\end{table}

\end{document}